# Structural Control of Metamaterial Oscillator Strength and Electric Field Enhancement at Terahertz Frequencies

G. R. Keiser[1*], H. R. Seren[2], A.C. Strikwerda[1,3], X. Zhang[2], and R. D. Averitt[1,4]

[1]Boston University, Department of Physics, Boston, MA 02215, USA
[2]Boston University, Department of Mechanical Engineering, Boston, MA 02215, USA
[3]Technical University of Denmark, DTU Fotonik – Department of Photonics Engineering, Kgs. Lyngby, DK-2800, Denmark
[4] UC San Diego, Department of Physics, La Jolla, CA 92093, USA
*grkeiser@physics.bu.edu

**Abstract: The design of artificial nonlinear materials requires control over internal resonant charge densities and local electric field distributions. We present a MM design with a structurally controllable oscillator strength and local electric field enhancement at terahertz frequencies. The MM consists of a split ring resonator (SRR) array stacked above an array of closed conducting rings. An in-plane, lateral shift of a half unit cell between the SRR and closed ring arrays results in an increase of the MM oscillator strength by a factor of 4 and a 40% change in the amplitude of the resonant electric field enhancement in the SRR capacitive gap. We use terahertz time-domain spectroscopy and numerical simulations to confirm our results. We show that the observed electromagnetic response in this MM is the result of image charges and currents induced in the closed rings by the SRR.**

Over the past decade, metamaterials (MM) research has produced innumerable examples and applications of artificially engineered index of refraction and impedance across the electromagnetic spectrum. Indeed, the onset of experimental MM research gave new life to long neglected theoretical speculation into optical behavior not observed in natural materials [1]. This renewed interest quickly bore experimental fruit, of which negative index materials [2,3], electromagnetic cloaking[4], perfect lensing [5,6], sub-diffraction imaging[7], and perfect absorption[8,9] are only a few paramount examples.

Though quite astounding in their own right, these examples form only the tip of the MM iceberg. MMs are useful not only for engineering far-field optical behavior like that discussed above but also allow for the design and control over microscopic charge and current distributions. These near-field properties form the cornerstone for a second category of MM research, specifically based around engineering local fields inside a MM unit cell. For instance, the on-resonance charge distributions in the capacitive gaps of a split ring resonator (SRR) MM [10,11] create localized regions inside the MM where the electric field of incident resonant radiation is greatly enhanced. These regions of electric field enhancement (EFE), when combined with complex material substrates and high intensity radiation sources, allows for the creation of nonlinear MM responses [12-15] and MM applications in high field, nonlinear spectroscopy [16].

Promising results and applications for MMs exist on all fronts [17,18]. However, the static nature and limited functional bandwidth of most MM designs present large challenges for MM engineering. Many applications in both the near-field and far-field regimes require broadband functionality and dynamic control of MM optical properties.

Fortunately, multiple approaches exist for altering and tuning a MM's electromagnetic behavior[12,19,20] or producing high functional bandwidths[21]. One promising approach involves altering the MM's unit cell geometry in order to change the near field interactions between different inclusions inside the MM[22]. Through these changes in near field interactions, structural changes to the size, shape, or relative position of inclusions within a MM can greatly alter the MM's near field or far field properties. The result is MM designs that promise high tunability [23,24] and more complex electromagnetic behavior [25]. Additionally, recent advances in MEMS actuation technology are now making such structural tuning a dynamic, real-time, and reversible process[26-28].

In this paper, we present a dual layer terahertz MM with structurally controllable oscillator strength and EFE. Our MM design consists of an SRR array placed above an array of closed conducting rings (See Fig. 1). An in-plane lateral displacement between the SRR and closed ring arrays results in an increase to the MM oscillator strength by a factor of 4 and a 40% change in the on-resonance EFE in the capacitive gaps of the SRRs. We present terahertz time domain spectroscopy (THz-TDS) measurements and numerical simulations to confirm these results. We show that the observed electromagnetic response in this MM is the result of image charges and currents induced in the closed rings by the SRR.

Photographs and a dimensioned schematic of our MM are presented in Fig. 1. The unit cell of the MM consists of a square SRR placed above a closed metallic ring (CR) of equal size. The SRR and CR are separated by a 5μm polyimide substrate ($\varepsilon_r$ = 2.88, loss-tangent tan($\delta$) = 0.0313). Both rings are then covered with a 5μm polyimide superstrate. The unit-cell periodicity is $P$=60μm, SRR and CR side-length $L_o$=40μm, metallization width $w$=11μm, and SRR capacitive gap width $g$=2μm. The lateral shift ($L_{shift}$) between the SRR and CR varies between samples from 0μm to 30μm. The dimensions are such that a 30μm shift is equal to a shift of half a unit cell. All samples were fabricated using conventional photolithography as described in detail in Ref.[24].

THz-TDS measurements were performed to characterize the MM response. The broadband THz pulse, with electric field polarized across the SRR capacitive gap as in Fig. 1(b), excited the SRR via the THz electric field. The resulting transmission spectra are shown in Fig. 2(a). A transmission minimum, corresponding to the SRR

LC resonance is distinctly visible in the data. The resonance transmission minimum shows a strong dependence on $L_{shift}$, decreasing by ~50% from $L_{shift}$=0μm to $L_{shift}$=30μm. Variations in $L_{shift}$ also cause small changes in the center frequency of the resonance. This is due to changes in the local environment of the SRR (and thus the lumped capacitance and inductance of the MM) as the CR is moved away from the SRR. The large frequency shifts seen in similar MM structures, i.e. the shifted broadside coupled SRR (BC-SRR)[23, 24], are not seen in this case because the CR is not resonant in the bandwidth of interest and no mode hybridization occurs[29].

To further investigate the $L_{shift}$ dependent MM response we performed numerical simulations of the SRR/CR structure in CST Microwave Studio. Simulated transmission spectra are presented alongside the experimental results in Fig. 2(b). The excellent correspondence between simulation and experiment allows for a more detailed analysis of the MM's internal behavior. Combining the simulated data with established parameter extraction techniques [30] reveals that the shift dependent transmission minimum corresponds to a changing MM oscillator strength. The extracted real part of the MM permittivity is shown in Fig. 3(a) and is well described by the real-part of a Lorentzian function as shown in Eq. 1, where $\varepsilon_\infty$ is the high frequency permittivity, $\omega_o$ is the resonance frequency, $\gamma$ is the damping frequency, and F is the oscillator strength.

$$\varepsilon(\omega) = \varepsilon_\infty - \frac{F\omega_o^2}{\omega_o^2 + i\gamma\omega + \omega^2} \qquad (1)$$

The on resonance maximum of the Lorentzian permittivity curves depends strongly on $L_{shift}$, suggesting that F depends on the relative lateral displacement between the SRR and CR. The MM also has a magnetic response due to the inherent bianisotropy of the SRR. [11] However, this response is much weaker than the electric response of the MM since the incident THz radiation does not couple to the SRR through the magnetic field in this experimental geometry.

The EFE inside the SRR capacitive gap is also strongly dependent on $L_{shift}$. Fig. 3(b) plots the simulated EFE at the center of the SRR capacitive gap vs. frequency, calculated using a local field probe in the CST transient solver, for varying values of $L_{shift}$. The on-resonance EFE increases considerably, changing by ~40%, as $L_{shift}$ increases from 0μm to 30μm.

Notably, while the CR does not have a resonance in the bandwidth of interest, currents are still induced in the CR through interactions with the SRR. A measure of this CR current, computed by integrating the current flow through a cross-section of one of the CR legs, is plotted in Fig. 3(c). Care must be taken when interpreting this scalar measure of a non-uniform vector current distribution. In general, the quantitative values of the currents in Fig. 3(c) will depend on the position of the integration plane in the CR legs. Yet, this current measure can still provide two qualitative insights into the origin of the MM's shift dependent properties. Specifically, it is clear that the CR's induced current resonates at the SRR LC resonance frequency and decreases in magnitude with increasing $L_{shift}$. Fig. 3(d) plots both the peak value of the CR scalar current and the magnitude of the MM oscillator strength, computed by fitting the extracted MM permittivity to Eq. 1, vs. $L_{shift}$. In fact, the oscillator strength increases by roughly a factor of 4 as the closed ring current decreases following a roughly inverse trend.

The behavior of the induced CR current suggests that interactions between the SRR and CR are responsible for the $L_{shift}$ dependent properties of this MM. A schematic model of the interactions between the SRR and CR is presented in Fig. 4(a) and 4(b). Consider the case for small values of $L_{shift}$. The CR sits directly above the SRR, as shown in Fig. 4(a). An incident terahertz electric field, polarized across the SRR capacitive gap will excite circulating currents in the SRR, creating resonant electric and magnetic dipole moments in the SRR. For electrical excitation, as in this experiment, the oscillator strength and in-gap EFE of the MM are both directly proportional to the strength of the net induced electric dipole moment, $P_{net}$, within the MM unit cell.

The CR affects the oscillator strength and EFE by acting as a plane for image charges and currents. The SRR resonant electric dipole moment, labeled $P_{SRR}$ in Fig. 4, induces image charges in the CR. These resonant image charges form an image dipole moment, $P_{Img}$, pointing opposite to $P_{SRR}$. Additionally, the oscillating image charges induced in the CR result in associated image currents in the CR that oscillate in the opposite direction as the currents in the SRR. This is the current discussed above in Fig. 3(c) and (d). Figure 5 shows the simulated resonant current distributions in the CR for different values of $L_{shift}$ compared to the resonant current distribution in the SRR. The "mirror" effect of the CR can clearly be seen as $L_{shift}$ is increased from 0 μm in Fig. 5(b) to 30 μm in Fig. 5(e).

Since $P_{SRR}$ and $P_{Img}$ act in opposite directions, the magnitude of the net MM dipole moment is determined by $P_{Net} = P_{SRR} - P_{Img}$. When the CR is shifted away from the SRR, as depicted in Fig. 4(b), the SRR is no longer directly above the CR image plane. Thus the dipole moment is no longer completely reflected in the CR, resulting in a smaller, $P_{Img}$. Thus, increasing $L_{shift}$ reduces $P_{Img}$ resulting in an increased $P_{Net}$ which, in turn, increases both the MM oscillator strength and the in-gap EFE.

It is important to note that though the SRR interacts with the CR to induce image charges and currents in the CR, the SRR is not coupling to one of the CR resonant modes. The CR has two resonant modes in the vicinity of the SRR's LC resonance.[31] The first is a static, uniform current mode at 0THz. The second mode is the dipole mode at ~2.5THz for the dimensions used in this study. The lack of a capacitive gap means that the CR does not have a LC resonant mode. The current distributions in Fig. 5 show that the CR current densities do not represent either of the possible modes of the CR, but instead image the resonant currents of the SRR. Additionally, if the effects discussed above were the result of mode coupling[29] between the SRR LC mode and a CR mode, a large, unidirectional frequency shift in the MM resonance mode would be present in the data which is not the case.

In summary, we have demonstrated THz metamaterials with structurally controllable oscillator strength and electric field enhancement at THz frequencies. Using THz-TDS and numerical simulations we investigated the electromagnetic response of a layered MM composed of an SRR array above an array of closed conducting rings. Image charges and currents induced in the closed rings via interactions with the SRR lower the oscillator strength of the SRR resonance. Laterally shifting the closed ring array relative to the SRR array reduces the reduces the image currents induced in the closed ring array, resulting in a increase to the oscillator strength of the MM by a factor of 4 and a ~40% change in the magnitude of the SRR's in-gap electric field enhancement. The growing importance and influence of electric field enhancement and other MM near field properties cannot be overstated, especially in nonlinear MMs and MM spectroscopic applications. This work provides the initial steps towards dynamic user control and tuning of these important MM properties.

The authors acknowledge support from DTRA C&B Technologies Directorate administered through a subcontract from ARL, and from NSF under contract number ECCS 1309835 and AFOSR under contract number FA9550-09-1-0708. The authors would also like to thank Logan Chieffo for several insightful discussions regarding simulation techniques and the Photonics Center at Boston University for technical support throughout this project.

**Figures**

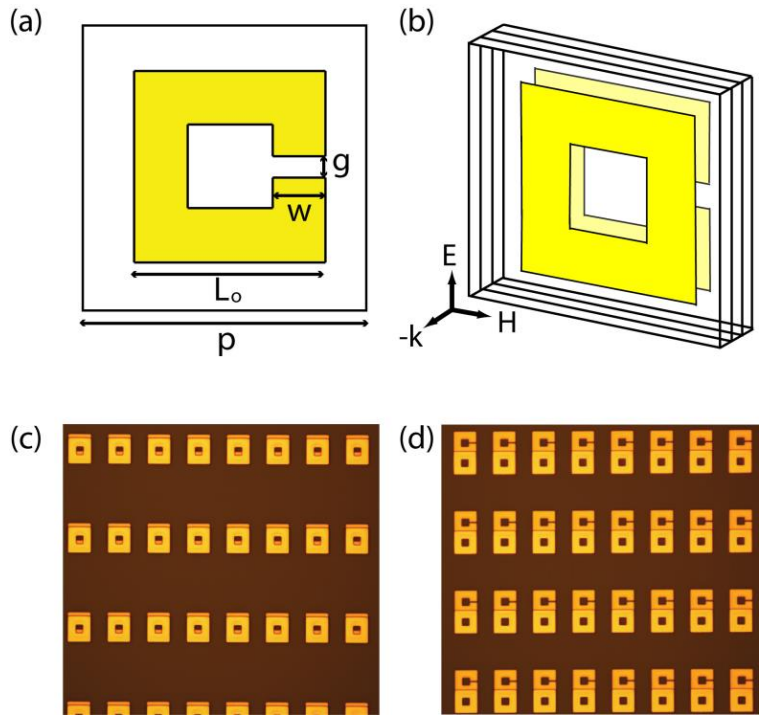

**Figure 1:** (a) SRR schematic with dimensions. (b) Perspective representation of the unit cell of a stacked SRR/CR metamaterial including the excitation direction and polarization. (c) and (d) Top-down photographs of SRR/CR MM samples with varying lateral displacement, i.e. varying $L_{shift}$, between the SRR and CR arrays.

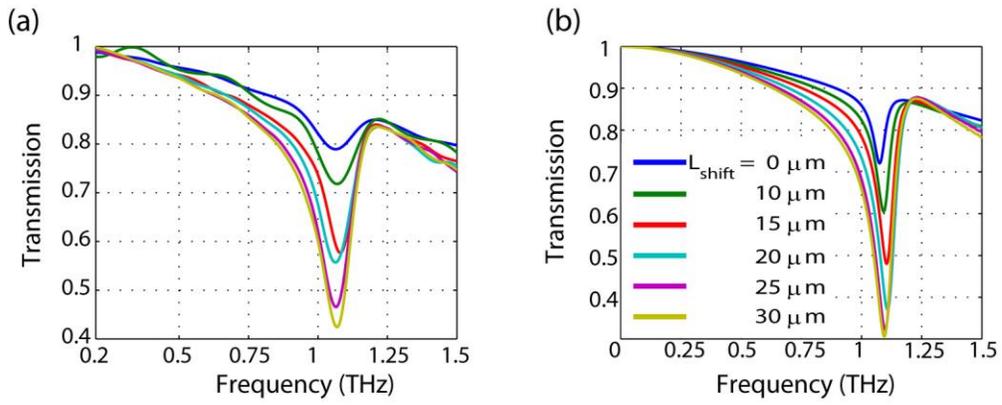

**Figure 2:** (a) Experimental THz-TDS transmission spectra for the SRR/CR MM for varying values of $L_{shift}$. (b) Simulated transmission spectra for the SRR/CR MM obtained via CST Microwave Studio.

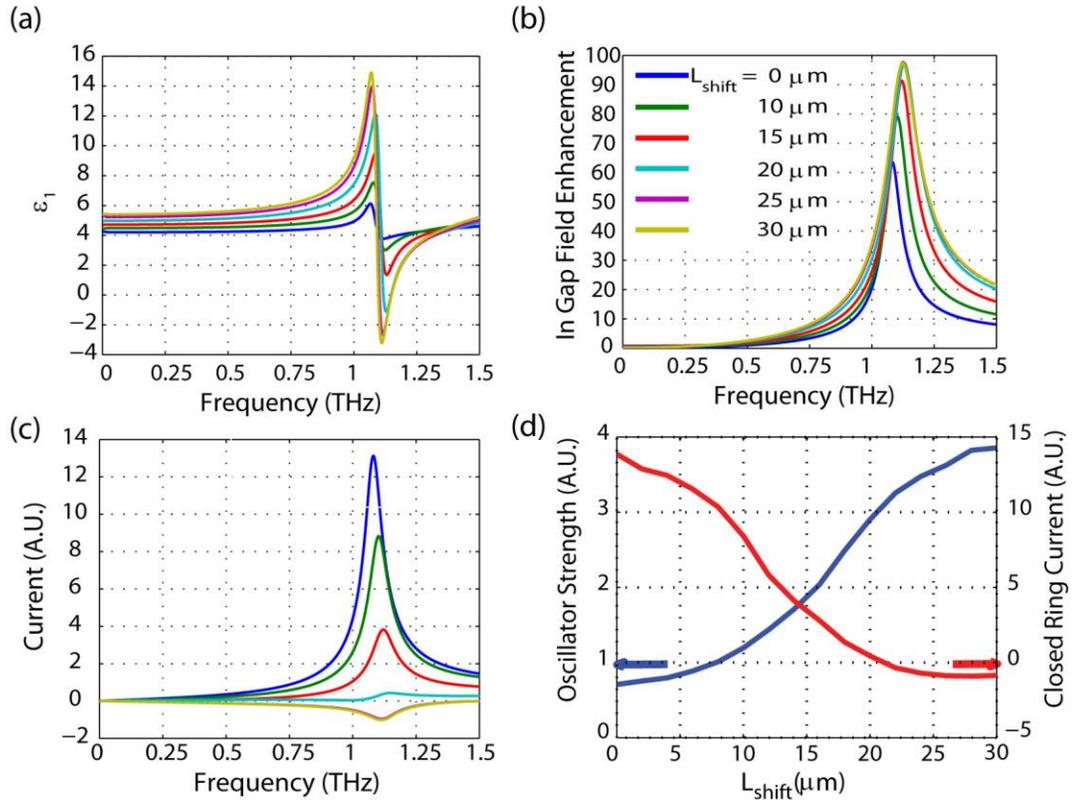

**Figure 3:** Extended analysis of the SRR/CR response via simulation and parameter extraction techniques. (a) Real part of the extracted permittivity of the SRR/CR MM for varying values of $L_{shift}$. (b) Simulated resonant electric field enhancement (EFE) in the SRR capacitive gap for varying values of $L_{shift}$. The EFE increases with increasing $L_{shift}$. (c) Magnitude of the induced current in the CR for varying values of $L_{shift}$. The CR current decreases with increasing $L_{shift}$. (d) Comparison of MM oscillator strength and CR current vs. $L_{shift}$.

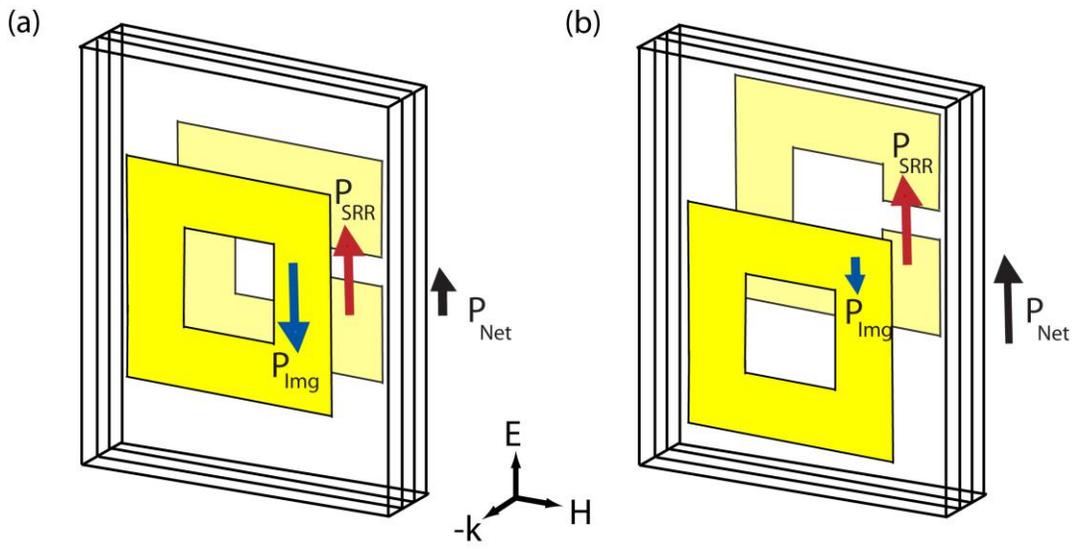

**Figure 4:** Coupling model for interactions between the SRR and the CR. (a) The SRR's resonant electric dipole moment, $P_{SRR}$, induces an image dipole $P_{Img}$ in the CR. These dipole moments are nearly equal in magnitude and opposite in direction for $L_{shift} = 0$ μm, resulting in a small $P_{Net}$. (b) Shifting the CR away from the SRR decreases the coupling between the SRR and CR, reducing the size of $P_{Img}$, which increases $P_{Net}$. A larger $P_{Net}$ leads to a larger oscillator strength and EFE.

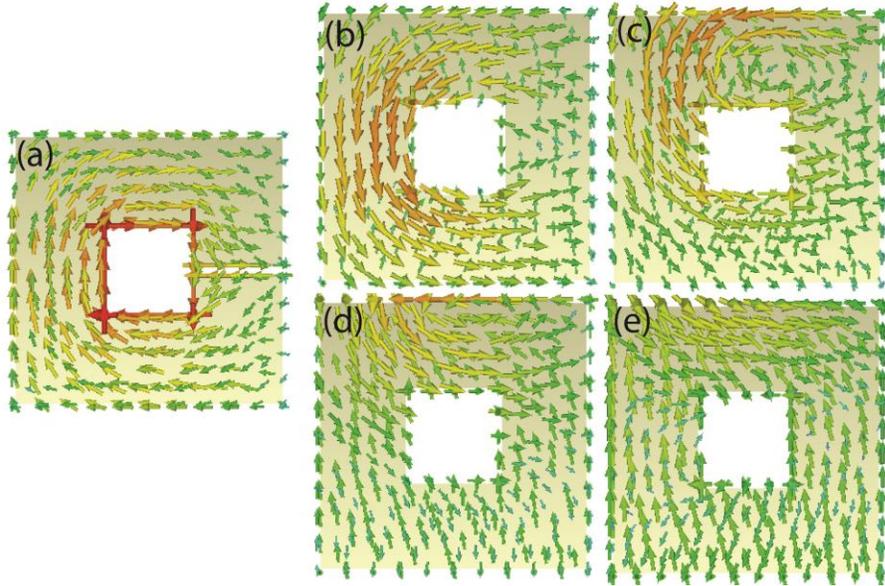

**Figure 5:** Image currents induced in the CR as a function of shift. (a) SRR LC current mode (b) CR image currents for $L_{shift}$ = 0 μm, (c) $L_{shift}$ = 10 μm (d) $L_{shift}$ = 20 μm, and (e) $L_{shift}$ = 30 μm. As the CR is shifted away from the SRR, the CR current mode remains centered at the same position, showing that the CR is acting as an image plane and mirroring the currents in the SRR.